\documentclass[aps,pre,twocolumn]{revtex4}
\usepackage{amsmath,bm,epsfig}
\usepackage{amssymb}
\newcommand{\B}[1]{{\bm{#1}}}
\newcommand{\C}[1]{{\mathcal{#1}}}    
\newcommand{\Hes}{\B{\C H}}
\newcommand{\ud}{\mathrm{d}}

\begin{document}

\title{Statistical Physics of Pure Barkhausen Noise}
\author{H. George E. Hentschel$^{1,2}$, Valery Iliyn$^2$, Itamar Procaccia$^2$ and Bhaskar Sen Gupta$^2$     }
\affiliation{$^1$Dept. of Physics, Emory University, Atlanta Ga.\\$^2$Dept. of Chemical Physics, The Weizmann Institute of Science, Rehovot 76100, Israel}


\begin{abstract}
We discuss a model metallic glass in which Barkhausen Noise can be studied in exquisite detail, free of thermal effects and of the rate of ramping of the magnetic field. The mechanism of the jumps in magnetic moment that cause the Barkhausen Noise can be fully understood as consecutive instabilities where an eigenvalue of the Hessian matrix hits zero, leading to a magnetization jump $\Delta m$ which is simultaneous with a stress and energy changes $\Delta \sigma$ and $\Delta U$ respectively. Contrary to common belief we find no ``movements of magnetic domain boundaries" across pinning sites, no fractal domains, no self-organized criticality and no exact scaling behaviour. We present a careful numerical analysis of the statistical properties of the phenomenon, and show that with every care taken this analysis is tricky, and easily misleading. Without a guiding theory it is almost impossible to get the right answer for the statistics of Barkhausen Noise. We therefore present an analytic theory, showing that the probability distribution function (pdf) of Barkhausen Noise is not a power law times an exponential cutoff.
\end{abstract}
\maketitle
\section{Introduction}

Barkhausen Noise which was discovered in 1919 \cite{19Bar}; it is a well known and frequently studied physical phenomenon. It is manifested as a series of jumps in the magnetization of a ferromagnetic sample when subjected to varying external magnetic field. Early reviews of the phenomenon and its implications can be found in Refs. \cite{69Bit, 76Bit,76MS}. The phenomenon has practical importance for magnetic recordings \cite{92BZ} and for noninvasive material characterization \cite{94Sip}. Various approaches to the statistics of Barkhausen Noise were critically discussed in Ref. \cite{96SBMS}; these approaches include assuming that Barkhausen Noise is due to domain-wall motion \cite{09DMW}, to Self-Organized Criticality \cite{91CM}, or to plain old critical phenomena \cite{95PDS}. Some authors proposed that Barkhausen Noise presents universal behavior (see for example \cite{01SDM}) whereas others argued against universality (see for example \cite{96Tad}). Indeed,
Barkhausen Noise appears to be a very complex physical phenomenon with many different appearances. Its character may depend on the type of ferromagnetic specimen under
study, the character of the disorder in the material, the external field driving rate, thermal effects, strength of the demagnetization fields, and other experimental details.

\begin{figure}
\vskip 0.5 cm
\includegraphics[scale = 0.32]{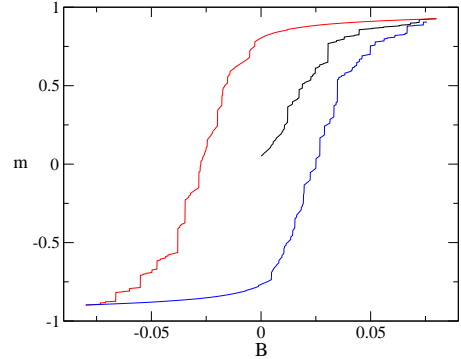}
\caption{A typical dependence of the magnetization $m$ on the magnetic field $B$, showing the initial increase from the freshly quenched state, and then the well known hysteresis curve. The discontinuous jumps in $m$ are the source of the Barkhausen noise whose statistics is the subject of this paper.}
\label{Bark}
\end{figure}

It is quite remarkable that although many of the experimental realization of Barkhausen Noise are obtained using metallic glasses as a medium, in fact none of the theoretical models presented in the literature attempted to approximate the physics of metallic glasses.
The aim of this paper is to close this gap, to study Barkhausen Noise in a model metallic glass that respects the glassy randomness of the materials and their magnetic properties. The model used by us is presented in detail in Sect. \ref{model}, but in Fig. \ref{Bark} we
present the main result of the numerical simulations of this model in the form of the classical hysteresis curve for the magnetization $\B m$ as a function of the external field $\B B$. Starting from the freshly quenched glass at zero field the magnetization increases with increasing the external field until saturation, at which point the magnetic field is reduced and then inverted until the opposite saturation. Finally the magnetic field is increased again. The magnetization curve has smooth sections punctuated by discontinuities whose size and distribution will be the focus of this paper. Using our model we can collect enormous amounts of data which allow us to study the phenomenon at exquisite detail. To reach the fundamental nature of the Barkhausen Noise we run all our simulations at zero temperature and with a quasi-static change of external magnetic field. We are thus free of thermal effects and of rate or ramping effects. We can determine precisely the nature of the discontinuities in the magnetization and show that at least in this model they stem from plastic events where the discontinuities in the magnetization are simultaneous with discontinuities in energy and in stress. We will show that here Barkhausen Noise has nothing to do with the movement of domain walls nor with Self Organized Criticality nor with fractal domains nor with thermodynamic criticality. Finally we will provide an analytic theory for the probability distribution function (pdf) of Barkhausen Noise, demonstrating explicitly the lack of exact scaling behavior. The analytic answer for the pdf of the jumps $\Delta m$ in magnetization is
\begin{equation}
\label{final}
P(\Delta m)=\frac{\exp(-A \Delta m ) }{\Delta m} f(\Delta m) \ ,
\end{equation}
where the exponential decay rate $A$ is analytically computed. The function $f(\Delta m)$ is evaluated, it is neither a power nor an exponent, and
it destroys the usually assumed form of the statistics of Barkhausen Noise.

In Sect. \ref{model} we present our model which was employed recently to study the interesting cross-effects between mechanics and magnetism in magnetic metallic glasses. The following section \ref{nature} discusses the mechanism for the jumps in magnetization that occur upon ramping the external magnetic field.  In Sect. \ref{statistics} we present the analysis of the statistics of Barkhausen Noise. We show that log-log plots of the distribution functions can provide terribly misleading results; getting more believable results requires the use of cumulative statistics and maximum-likelihood methods. But also these methods fail to reach the truth, leading us to believe that the pdf of $\Delta m$ is given by a power law times and exponential cutoff. In Sect. \ref{theory} we present a theory of Barkhausen Noise in our system and find the analytic form Eq. (\ref{final}). We demonstrate quantitative agreement between theory and simulations. Finally in Sect. \ref{summary} we offer a summary and a discussion of the results presented in this paper.

\section{The Model}
\label{model}

Our model Hamiltonian is in the spirit of the Harris, Plischke and Zuckerman (HPZ) Hamiltonian \cite{73HPZ}
but with a number of important modifications to conform with the physics of amorphous magnetic solids \cite{12HIP}. One important difference is that our particles are not pinned to a lattice.  We write the Hamiltonian as
\begin{equation}
\label{umech}
U(\{\B r_i\},\{\B S_i\}) = U_{\rm mech}(\{\B r_i\}) + U_{\rm mag}(\{\B r_i\},\{\B S_i\})\ ,
\end{equation}
where $\{\B r_i\}_{i=1}^N$ are the 2-D positions of $N$ particles in an area $L^2$ and $\B S_i$ are spin variables. The mechanical part $U_{\rm mech}$ is chosen to represent a glassy material with a binary mixture of 65\% particles A and 35\% particles B,
with Lennard-Jones potentials having a minimum at positions $\sigma_{AA}=1.17557$, $\sigma_{AB}=1.0$ and $\sigma_{BB}=0.618034$ for the corresponding interacting particles \cite{09BSPK}. These values are chosen to guarantee good glass formation and avoidance of crystallization. The energy parameters chosen are $\epsilon_{AA}=\epsilon_{BB}=0.5$
$\epsilon_{AB}=1.0$, in units for which the Boltzmann constant equals unity. All the potentials are truncated at distance 2.5$\sigma$ with two continuous derivatives. $N_A$ particles A carry spins $\B S_i$; the $N_B$ B particles are not magnetic. Of course $N_A+N_B= N$. We choose the spins $\B S_i$ to be classical $xy$ spins; the orientation of each spin is then given by an angle $\phi_i$ with respect to the direction of the external magnetic field which is along the $x$ axis.

The magnetic contribution to the potential energy takes the form \cite{12HIP}:
\begin{eqnarray}
&&U_{\rm mag}(\{\B r_i\}, \{\B S_i\}) = - \sum_{<ij>}J(r_{ij}) \cos{(\phi_i-\phi_j)}\nonumber\\&&-  \sum_i K_i\cos^2{(\phi_i-\theta_i(\{\B r_i\}))}-  \mu_A B \sum_i \cos{(\phi_i)} \ .
\label{magU}
\end{eqnarray}
Here $r_{ij}\equiv |\B r_i-\B r_j|$ and the sums are only over the A particles that carry spins. For a discussion of the physical significance of each term the reader is referred to Ref.~\cite{12HIP}. It is important however to stress that in our model (in contradistinction with the HPZ Hamiltonian \cite{73HPZ} and also with the Random Field Ising Model \cite{95PDS}), the exchange parameter $J(\B r_{ij})$ is a function of a changing inter-particle position (either due affine motions induced
by an external strain or an external magnetic field or due to non-affine particle displacements, and see below). Thus randomness in the exchange interaction is coming from the random positions $\{\B r_i\}$, whereas the function $J(\B r_{ij})$ is not random. We choose for concreteness the monotonically decreasing form $J(x) =J_0 f(x)$ where $f(x) \equiv \exp(-x^2/0.28)+H_0+H_2 x^2+H_4 x^4 $ with
$H_0=-5.51\times 10^{-8}\ ,H_2=1.68 \times 10^{-8}\ , H_4=-1.29 \times 10^{-9}$.
This choice cuts off $J(x)$ at $x=2.5$ with two smooth derivatives.  Note that we need to have at least two smooth derivatives in order to compute the Hessian matrix below. Finally, in our case $J_0=3$.

Another important difference with the HPZ model is that in our case
the local axis of anisotropy $\theta_i$ is {\em not} selected from a pre-determined distribution, but is determined by the local structure. In other words, in a crystalline solid the easy axis is determined by the symmetries of the lattice. In an amorphous solid the structure and the arrangement of particles changes from place to place, and we need to find the local easy axis by taking this arrangement into account. To this aim define  the matrix $\B T_i$:
\begin{equation}
T_i^{\alpha\beta} \equiv \sum_j J( r_{ij})  r_{ij}^\alpha r_{ij}^\beta/\sum_j J( r_{ij}) \ .
\end{equation}
Note that we sum over all the particles that are within the range of $J( r_{ij})$; this catches the arrangement of the local neighborhood of the $i$th particle. The matrix $\B T_i$ has two eigenvalues in 2-dimensions that we denote as $\kappa_{i,1}$ and $\kappa_{i,2}$, $\kappa_{i,1}\ge \kappa_{i,2}$. The eigenvector that belongs to the larger eigenvalue $\kappa_{i,1}$ is denoted by $\hat {\B n}$. The easy axis of anisotropy is given by $\theta_i\equiv \sin^{-1} (|\hat n_y|)$. Finally the coefficient $K_i$ which now changes from particle to particle is defined as
\begin{equation}
\label{KK}
K_i \equiv \tilde C[\sum_j J( r_{ij})]^2 (\kappa_{i,1}-\kappa_{i,2})^2\ ,~~ \tilde C= K_0/J_0\sigma^4_{AB} \ .
\end{equation}
The parameter $K_0$ determines the strength of this random local anisotropy term compared to other terms in the Hamiltonian. For most of the data shown below we chose $K_0=5.0$. The form given by Eq.~(\ref{KK}) ensures that for an isotropic distribution of particles $K_i=0$. Due to the glassy random nature of our material the direction $\theta_i$ is random. In fact we will assume below (as can be easily tested in the numerical simulations) that the angles $\theta_i$ are distributed randomly in the interval $[-\pi,\pi]$. It is important to note that external straining does NOT change this flat distribution and we will assert that the probability distribution $P(\theta_i)$ can be simply taken as
\begin{equation}
P(\theta_i)d\theta_i = \frac{d\theta_i}{2\pi} \ .
\label{ptheta}
\end{equation}
The last term in Eq. (\ref{magU}) is
the interaction with the external field $B$. We have chosen $\mu_A B$ in the range [-0.08,0.08]. At the two extreme values all the spins are aligned along the direction of $\B B$.

\section{The nature of the Barkhausen Noise}
\label{nature}

To simulate Barkhausen Noise with our model we first prepare a system with 2000 particles at constant volume and temperature $T=1.2$ with density $\rho=0.976$.  The system is equilibrated at this temperature using $10^5$ Monte Carlo sweeps. Next the system was cooled down to $T=0.6$ and equilibrated again using again $10^5$ Monte Carlo sweeps. Then the temperature was reduced by steps of $\Delta T=0.1$ down to $T=0.2$ with equilibration after every step. Finally the system was cooled down to $T=0.001$ by steps of $\Delta T=0.01$ and then $\Delta T=0.001$, equilibrating after every step. Subsequently the system is kept at temperature $T=0.001$ which is sufficiently low to eliminate any appreciable
thermal effects. At this point we begin to ramp the external magnetic field
in the $x$ direction in small steps of $\Delta B=10^{-4}$. After every such increase in magnetic field we minimize the energy by a conjugate gradient method. This quasi-static increase in magnetic field eliminates any effects of rate of ramping. We measure the magnetization $m$ defined as
\begin{equation}
m= \frac{1}{N_A} \sum_1^{N_A} \cos \phi_i \ .
\end{equation}
As seen in Fig. \ref{Bark} the magnetization starts at $m=0$ and increases upon increasing B until it saturates at $m=1$. At this point the magnetic field is reduced until the magnetization is saturated at $m=-1$. Finally the magnetic field is increased again to close a hysteresis loop. We can repeat this process many times, and in every cycle the smooth sections of the magnetization curve would be punctuated by discontinuities $\Delta m$ which occur at apparently random values of the external field $B$.

\begin{figure}
\includegraphics[scale = 0.35]{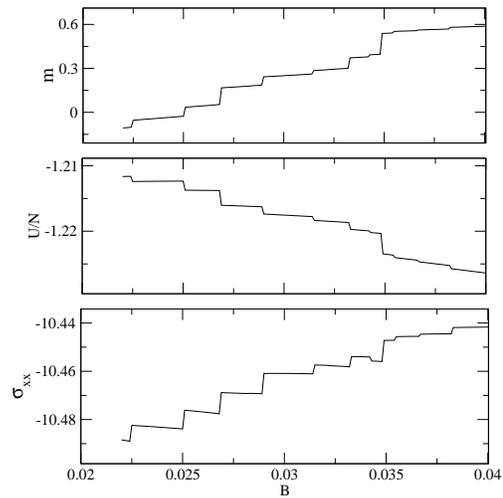}
\caption{The magnetization, the energy per particle and the stress component $\sigma_{xx}$ as a function of external magnetic field $B$. This figure demonstrates that all three quantities have discontinuities at the same value of $B$ where the system undergoes a plastic event with one of the eigenvalues of the Hessian matrix $\Hes$ hits zero, and the next figure.}
\label{changes}
\end{figure}

The focus of this section is on the physics underlying the discontinuities in the magnetization curve. In our case there is nothing mysterious about them. It is easy to see that the magnetization curve is smooth as long as the system is mechanically and magnetically stable. This is the case as long as the Hessian matrix $\Hes$ has only positive eigenvalues. In the present case $\Hes$ takes on the form \cite{12HIP}:
\begin{equation}
\label{Hesa}
\Hes =
\begin{pmatrix}
  \frac{\partial^2U}{\partial \B r_i\partial \B r_j} & \frac{\partial^2U}{\partial \B r_i\partial \phi_j} \\
  \frac{\partial^2U}{\partial \phi_i\partial \B r_i}  & \frac{\partial^2U}{\partial \phi_i\partial \phi_j} \
\end{pmatrix} \ .
\end{equation}
The system loses stability when at least one of the eigenvalues of $\Hes$ goes to zero. When this happens, there appears an instability that results in a discontinues change in stress, in energy and in magnetization. In Fig. \ref{changes} we show a typical blown up section of the energy, magnetization and stress curves as a function of $B$.
We see that the discontinuities appears simultaneously in alls the three quantities at the same values of $B$. These are irreversible plastic events that take the system from one minimum in the energy landscape through a saddle-node bifurcation to another minimum in the energy landscape where again all the eigenvalues of $\Hes$ are positive. In Ref. \cite{12HIP} we derived an exact equation for the dependence of any eigenvalue $\lambda_k$ on $B$ for a fixed external strain, which reads:
\begin{equation}
\frac{\partial \lambda_k}{\partial B}{\bf |}_{\gamma}   =   c^{(b)}_{kk} - \sum_\ell \frac{a^{(b)}_\ell [b^{(r)}_{kk\ell}+b^{(\phi )}_{kk\ell}]}{\lambda_\ell} .
\label{diff}
\end{equation}
The precise definition of all the coefficients is given explicitly in Ref. \cite{12HIP}. Generically, when one eigenvalue, say $\lambda_P$ approaches zero, all the other terms in Eq. (\ref{diff}) remain bounded, leading to the approximate equation
\begin{equation}
\frac{\partial \lambda_P}{\partial B}{\bf |}_{\gamma}\approx \frac{\text{Const.}}{\lambda_P} \ . \label{sqsing}
\end{equation}
In such generic situations the eigenvalue is expected to vanish following a square-root singularity, $\lambda_P \sim (B_p-B)^{1/2}$ where $B_p$ is the value of the external magnetic field where the eigenvalue vanishes. The reader should be aware of the fact that at some special values of $B$ it may happen that the coefficient Const in Eq. \ref{sqsing} vanishes at the
instability leading to a an exponent different from 1/2 \cite{13DHPS}. This non generic feature hardly changes the considerations of the present paper).
\begin{figure}
\includegraphics[scale = 0.35]{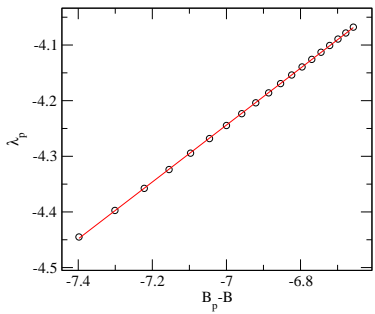}
\caption{The logarithm of the eigenvalue $\lambda_P$ that hits zero at $B_P$ as a function of the logarithm of $B_P-B$. The slope has a value of 1/2. }
\label{instab}
\end{figure}
In Fig. \ref{instab} we show a typical dependence of the eigenvalue $\lambda_P$ on $B$, where the square-root singularity is apparent. It is also interesting to examine what happens to the eigenfunctions $\B \Psi^{k}$ which are associated with the eigenvalues $\lambda_k$ as the instability is approached. The answer is that all the eigenfunctions of $\Hes$ are delocalized far from the instability, but the one eigenfunction $\B \Psi^{P}$ associated with $\lambda_P\to 0$ gets localized on $n\ll N$ particles. A typical projection of $\B \Psi^{P}$ close to the instability on the particles positions and on the spins is shown in the two panels of Fig. \ref{proj}.
\begin{figure}
\includegraphics[scale = 0.35]{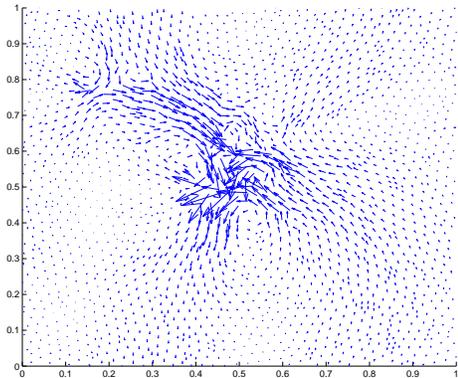}
\includegraphics[scale = 0.35]{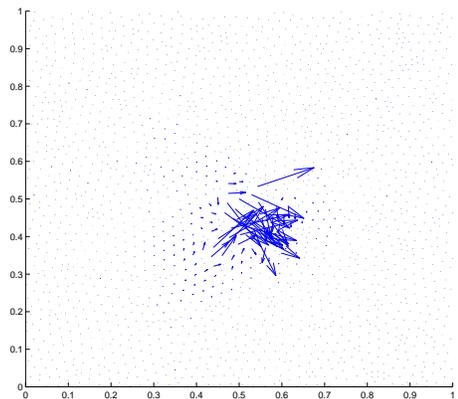}
\caption{The projection of the eigenfunction $\B \Psi^{P}$ associated with the eigenvalue $\lambda_P$ shown in Fig. \ref{instab} projected on the particles positions and on the spins in the upper and lower panels respectively. The upper panel shows a typical non-affine displacement field associated with a plastic event, having the quadrupolar structure of an Eshelby solution. The lower panel shows that the same event is associated with a co-local flip of spins, leading to the change $\delta m$ of the Barkhausen Noise.}
\label{proj}
\end{figure}
We see that the non-affine movement of the particles is very similar to the standard ``Eshelby like" quadrupolar event that is so typical to amorphous solids. The projection on the spin shows that a patch of spins had changed its orientation (magnetic flip of a domain). Note that the patch is compact, without any fractal or other esoteric characteristics that were associated with Barkhausen Noise in the past.  This is the nature of the event that is associated with the Barkhausen Noise in our case.
\section{Statistics of the Barkhausen Noise}
\label{statistics}

\subsection{Preliminaries}

\begin{figure}
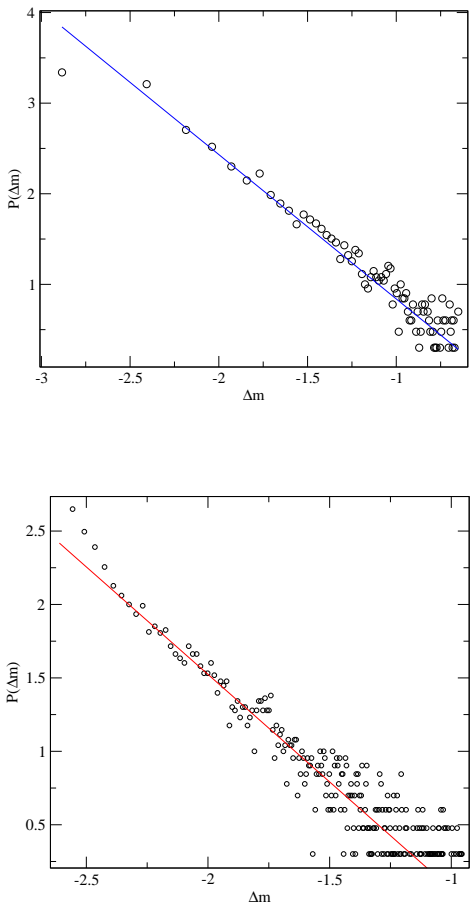

\includegraphics[scale = 0.35]{BarkFig5a.eps}
\vskip 1.2 cm
\includegraphics[scale = 0.36]{BarkFig5b.eps}
\caption{Demonstration of the inadequacy of log-log plots for characterizing the statistics of Barkhausen Noise. The two panels differ only the bin size employed to determine the probability to see a value of $\Delta m$. In the upper panel the bin size is 0.0026 and in the lower 0.00033. One sees an apparent scaling behavior $p(\delta m)\sim (\delta m)^{-\alpha}$ with two respective slopes $\alpha\approx 1.6$ in the upper panel and $\alpha \approx 1.46$ in the lower panel.
In between one can find a bin size to conform with $\alpha=1.5$ which is of course meaningless.  The conclusion is that we must avoid binning, and rather use cumulative distribution functions, see Eqs.~(\ref{defF}) and (\ref{comp}). }
\label{loglog}
\end{figure}

Typical statistics were accumulated from 50 hysteresis loops, where values of $\Delta m > (\Delta m)_{\rm min} =10^{-4}$ were carefully measured and stored. The largest values of $\Delta m$ found in this model are of the order of $10^{-1}$. We thus have three order of magnitude of $\Delta m$ allowing us to determine the statistics with satisfactory precision. One thing that one should NOT do is to bin the data and to plot log-log plots. In Fig. \ref{loglog} we show such typical plots for different bin size, to demonstrate that any power law can be justified by choosing the bin size. To avoid binning, we consider the cumulative distribution function. Defining the fundamental probability distribution function to see a value of $\Delta m$ such that $x\le \Delta m\le x+dx$ as $p(x) dx$, the cumulative function is defined as
\begin{equation}
F(y) \equiv \int_{(\Delta m)_{\rm min}}^y p(x) dx \ .
\label{defF}
\end{equation}
Associated with this function one also defines the complimentary cumulative distribution function as
\begin{equation}
F_c(y) =1-F(y) \ . \label{comp}
\end{equation}
\subsection{Analysis}

\begin{figure}
\includegraphics[scale = 0.35]{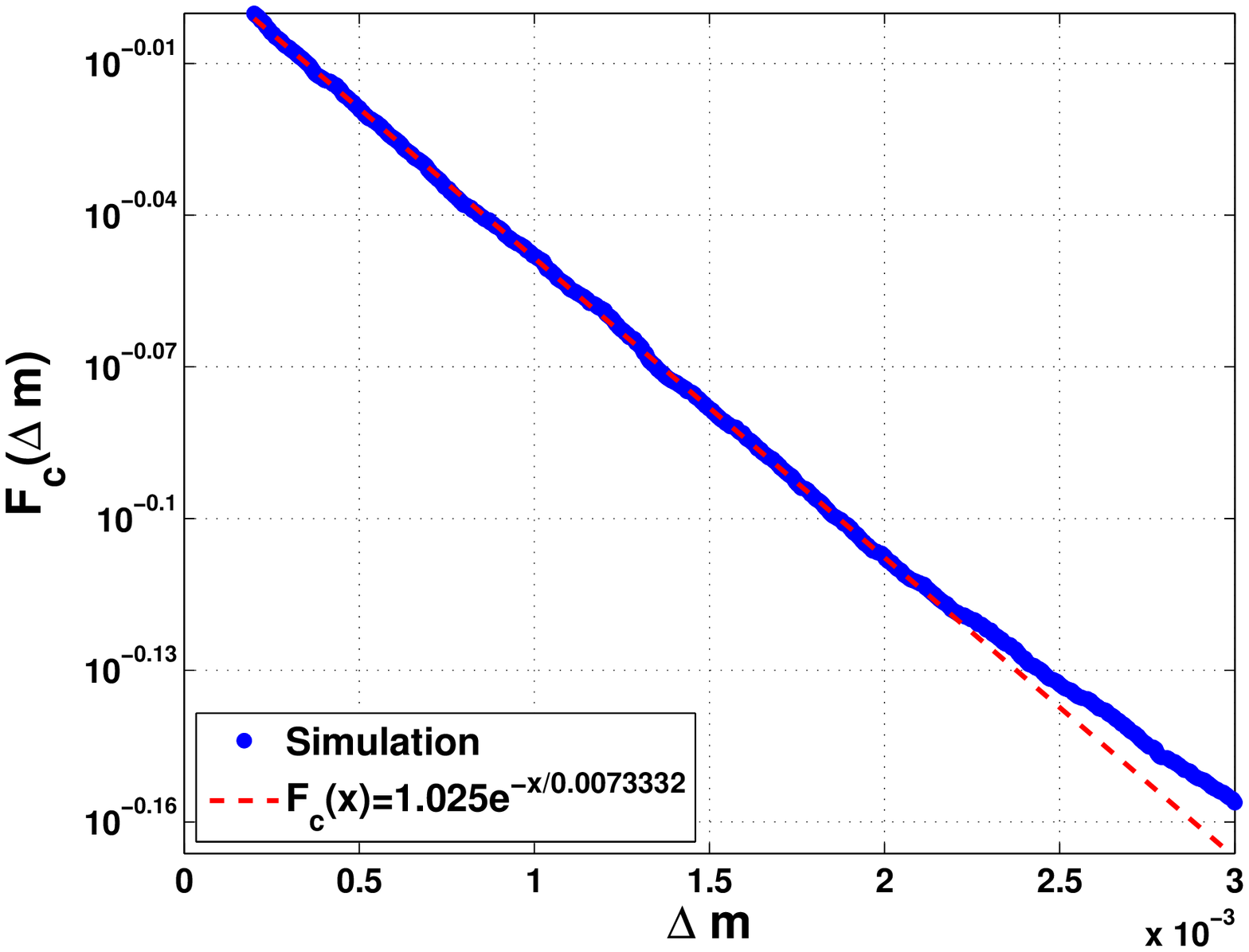}
\includegraphics[scale = 0.35]{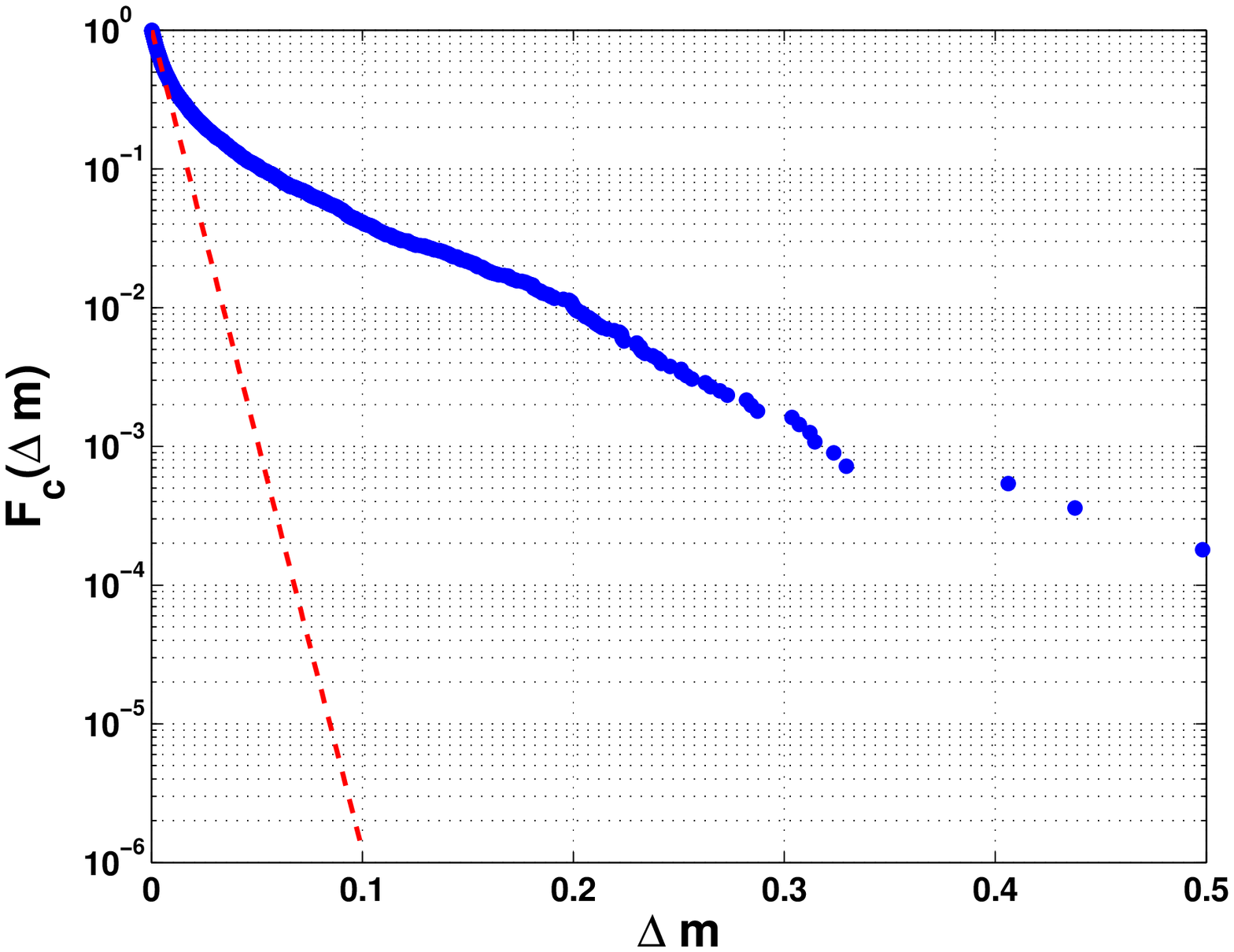}
\caption{The complimentary cumulative function $F_c(\Delta m)$ as a function of the upper limit $\Delta m$. In the upper panel one sees the change in behavior around $m=M_0=0.002$. In the lower panel the same function is presented in log-log plot, showing that for $\Delta m\le M_0$ and exponential function provides a very good fit. }
\label{smalldel}
\end{figure}

Analyzing the obtained data for $F_c(y)$ one immediately encounters a difficulty, i.e. that there is no single functional form that can be fitted to the data for all values of $y$. Very small jumps $\Delta m<M_0 \approx 0.002$ need to be analyzed separately from larger jumps. To see this we present in Fig. \ref{smalldel} the function $F_c(\Delta m)$ as a function of $\Delta m$ in log-linear and log-log plots. There is a clear change in behavior around $\Delta m =0.002$, such that below this value we can fit the data excellently well to an exponential function
\begin{equation}
F_c(\Delta m) \approx A \exp(-B~\Delta m)\ , \quad \text{for}~\Delta m \le 0.002 \ .
\end{equation}
For the model parameters reported above we find $A\approx 1.025$, $B\approx 136.37$. For values of $\Delta m> M_0$ the nature of the distribution changes qualitatively, and we need to analyze the data with a different value of $(\Delta m)_{\rm min}=0.002$ in Eq. (\ref{defF}). This unfortunately decreases our range of jumps $\Delta m$ to only two and half order of magnitude, but this is unavoidable in view of what is found.

\begin{figure}
\includegraphics[scale = 0.35]{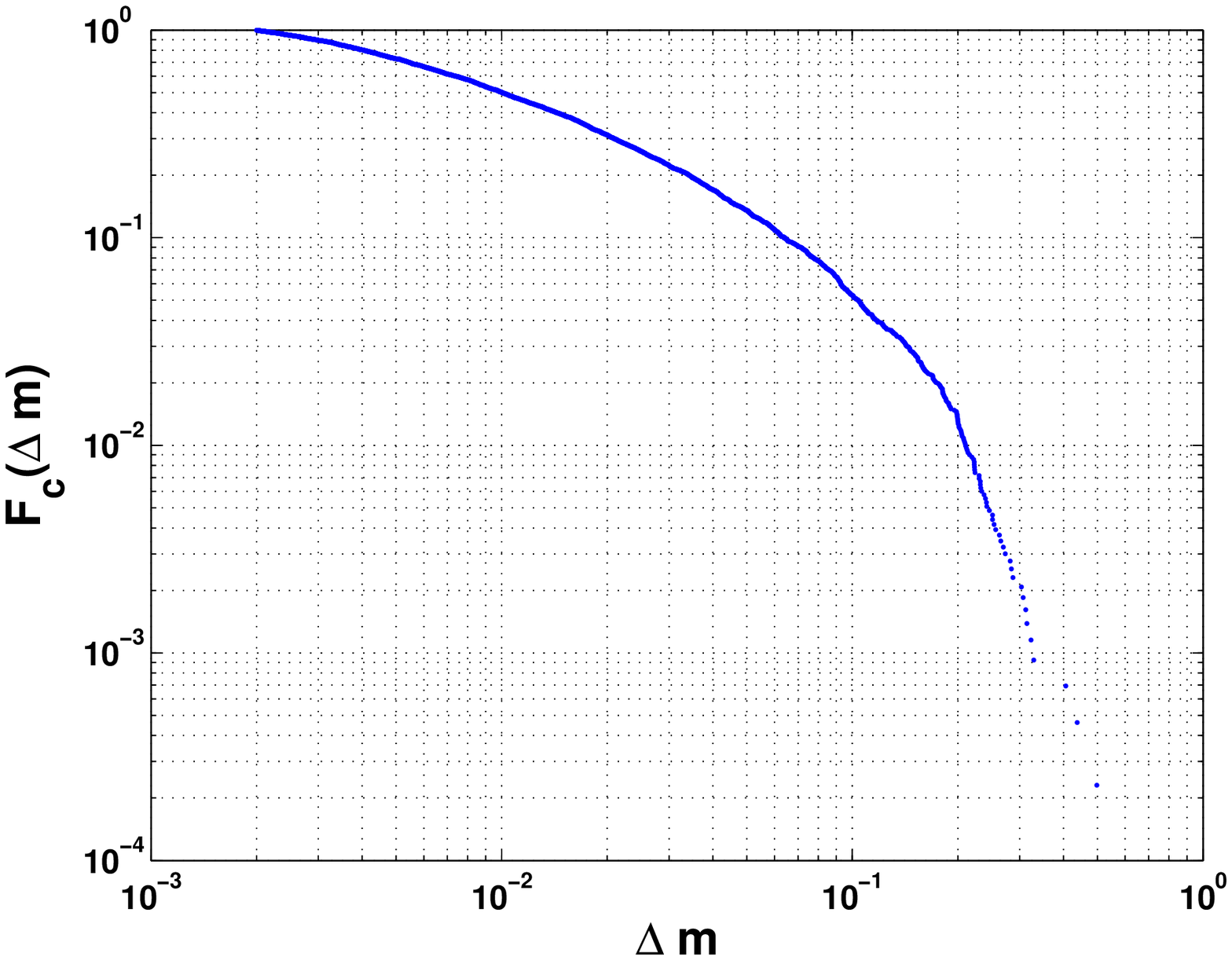}
\includegraphics[scale = 0.35]{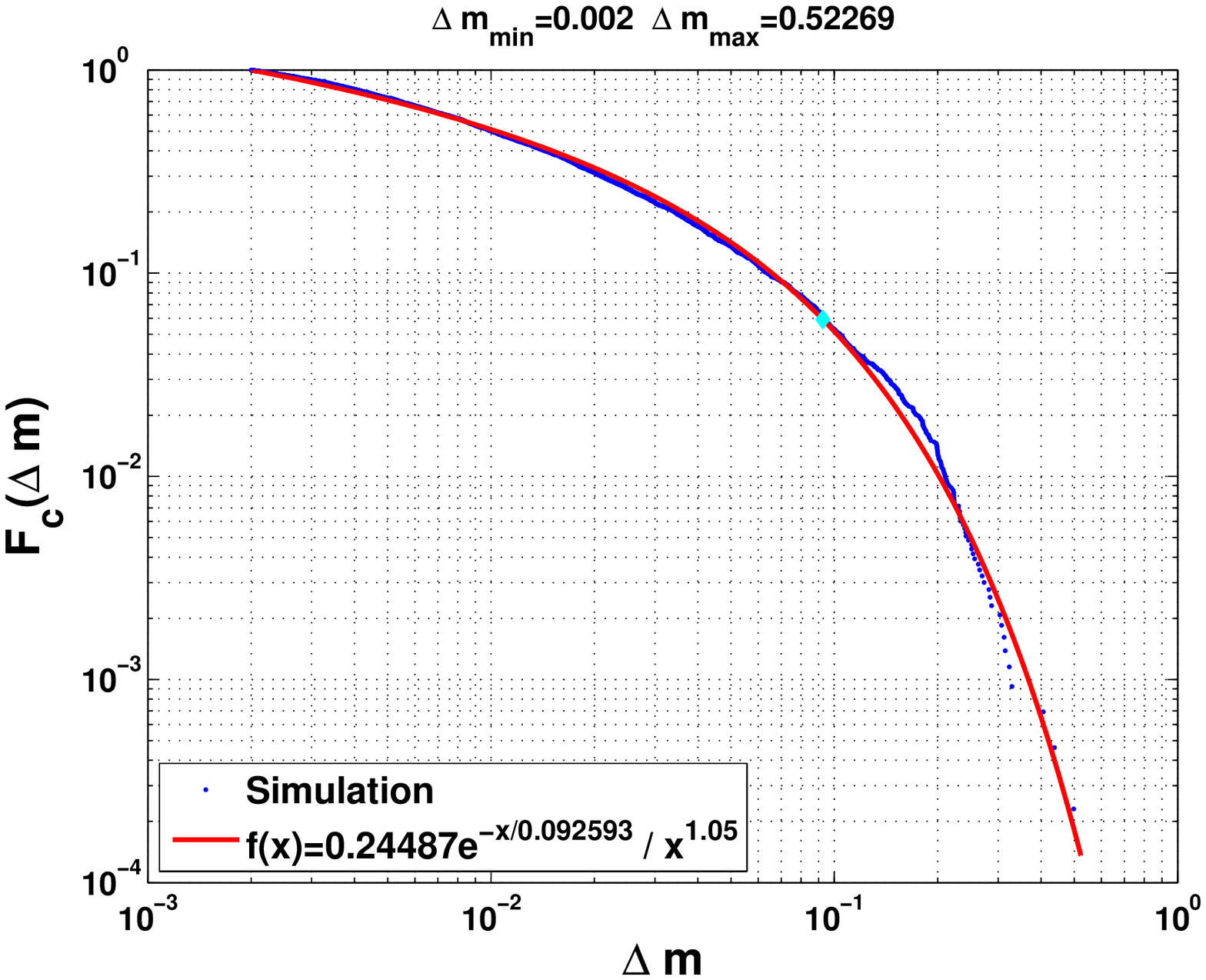}
\caption{Upper panel: The complimentary cumulative function $F_c(\Delta m)$ for $0.002\le \Delta m\le 0.52269$ in double logarithmic presentation. Lower panel: comparison between the data and the computed complimentary cumulative function resulting from integrating the fundamental pdf Eq. (\ref{pofx}). The agreement appear quite satisfactory. }
\label{comcum}
\end{figure}

Repeating the analysis of the complimentary cumulative distribution function in the range $0.002\le \Delta m \le 0.52269$ we obtain the function shown in log-log plot in Fig. \ref{comcum}. The present function appears to the result of a fundamental pdf $p(x)$ if the form
\begin{equation}
p(\Delta m) =C x^{-\alpha} \exp (-D \Delta m) \label{form}.
\end{equation}
To find the best values of the parameters $\alpha$ and $D$ (C is determined by normalization) we use the method of maximum likelihood (see appendix).
The best fit to the data is obtained as
\begin{equation}
p(\Delta m) =0.24487x^{-1.06} \exp (-10.8 \Delta m). \label{pofx}
\end{equation}
Having this trial function we can integrate it and compare with the  complimentary cumulative function that is generated by the data. This is done in the lower panel of Fig. \ref{comcum} with an apparent satisfactory agreement. We could be thus led to conclude that the pdf of Barkhausen Noise in our present model for values of $\Delta m>M_0=0.002$ is very well represented by a power law truncated with an exponential cutoff. The power law exponent has nothing to do with the sometime claimed universal value of 1.5. The latter can always be obtained from log-log plots with a well chosen binning, but is therefore irrelevant.

In fact, such a conclusion would be erroneous as well. To see the danger  under very sharp light we can repeat the very same analysis presented here for cumulative functions $F(y)$ but instead of using the minimal value
$M_{\rm min}=0.002$ we now use variable values $0.02\le M_{\rm min}=0.12$.
To our horror we find that for every value of $M_{\rm min}$ we can demonstrate equally good fit to a cumulative function which is derived
from a fundamental pdf function of the form of Eq. (\ref{form}) but with values of the exponent $\alpha$ ranging continuously from 1.06 to about 0.5. Of course, this is a strong warning that the correct underlying pdf is not of the form (\ref{form}) and that even with the care taken to avoid binning, we cannot guess the correct pdf. There is no escape, one must turn to theory in order to find the truth.

\section{Theory of Barkhausen Noise}
\label{theory}
\subsection{Magnetic domains}
\begin{figure}
\includegraphics[scale = 0.26]{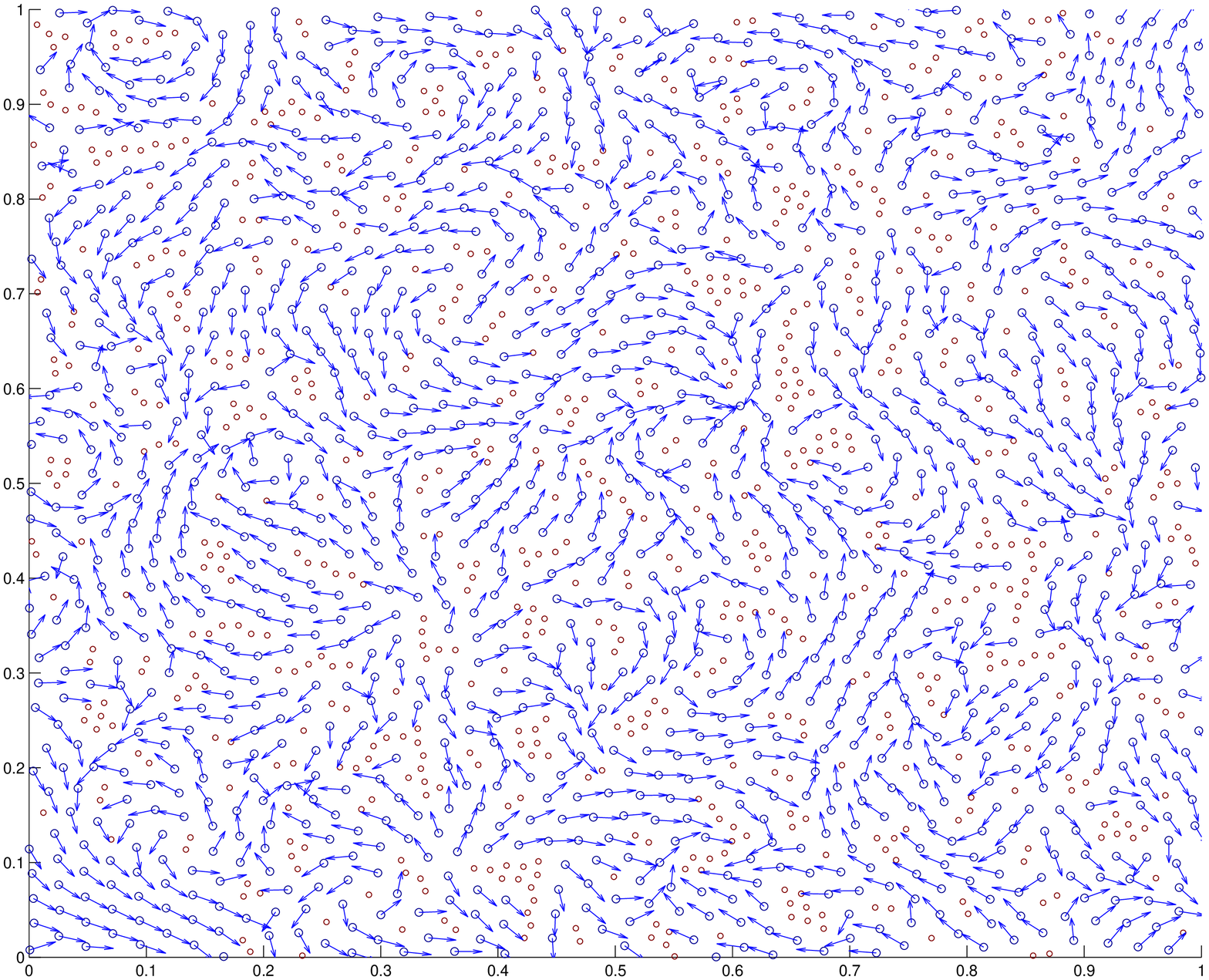}
\includegraphics[scale = 0.33]{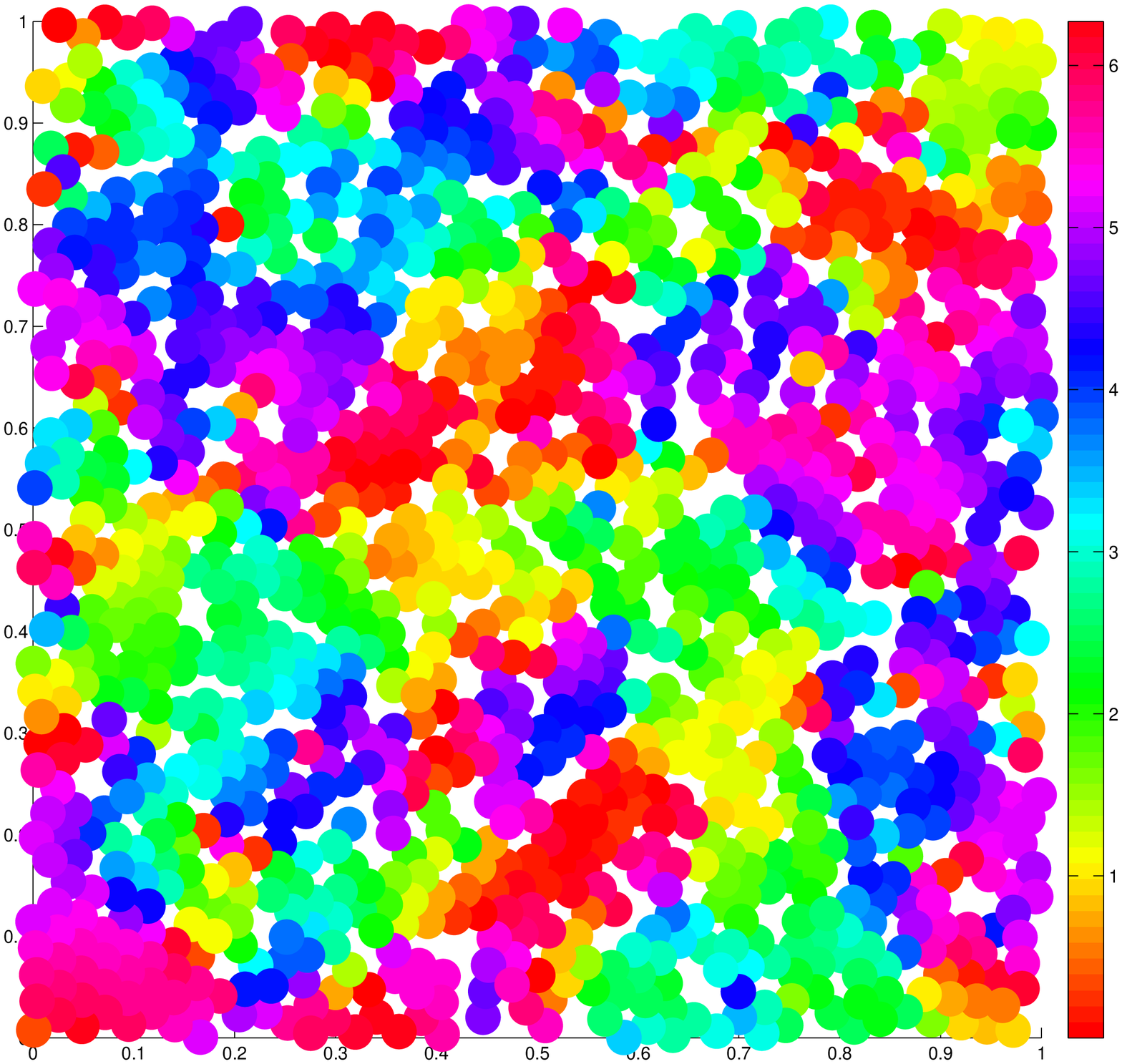}
\caption{Upper panel: A snapshot of the spin distribution in our magnetic glass. One can see the magnetic domain with bare eyes, but better after color coding as is shown in the lower panel. Lower panel: the same distribution of spins color coded according to orientation, see the color code on the right of the figure. }
\label{spins}
\end{figure}

In order to understand the Barkhausen Noise in the present model we must realize that our system at $B=0$ contains lots of magnetic domains in which the spins are pointing roughly in the same directions. A snapshot of the spin orientation in a typical realization of our magnetic glass is shown in Fig. \ref{spins} upper panel, with color coding in the lower panel. A given color means that there exists an average orientation of the spins in that domain, and below we will denote this average orientation as $\phi$, without the index $i$. Similarly, the average over $\theta_i$ in the domain will be denoted as $\theta$.
It becomes obvious that we can assume that the disordered spin distribution can be treated as a set of $\C N$ domains, such that there exist $\C N_s$ domains
consisting of $s\ge s_{min}$ quasi-ordered spins. Then we define $p_s$ to be the probability that an arbitrary chosen spin belongs to a domain of $s$ spins, with the normalization condition
\begin{equation}
\sum\limits_{s=s_{min}}^{s_{max}} p_s=1,
\label{norm}
\end{equation}
The mean domain spin value is fixed by
\begin{equation}
\sum\limits_{s=s_{min}}^{s_{max}} s p_s=\langle s\rangle.
\label{mean}
\end{equation}

In accordance with the principle of maximum entropy \cite{J57} to find the actual distribution $p_s$ we should maximize the information entropy
\begin{equation}
S=-\sum\limits_{s=s_{min}}^{s_{max}} p_s\ln p_s,
\label{entr}
\end{equation}
subject to the constrains defined by
Eq.~(\ref{norm}) and Eq.~(\ref{mean}). The standard method of   Lagrange multipliers \cite{J57} is detailed in Appendix \ref{maxent}
with the final result
\begin{equation}
p_s=\frac{e^{-(s-s_{min})/\langle s\rangle}}{\langle s\rangle}\ . \label{pofs}
\end{equation}

The fraction of domains that contains $s$ spins is defined by
\begin{equation}
F_s=\frac{\C N_s}{\sum\limits_{s=s_{min}}^{s_{max}} \C N_s}.
\label{Fr}
\end{equation}
The probability distribution function $p_s$ by the definition is given by
\begin{equation}
p_s=\frac{s\C N_s}{\C N}.
\label{pDef}
\end{equation}
Below it is more convenient to introduce a new variable related to the
magnetization $x=s/N_A$.
It follows from Eq.~(\ref{Fr}) and Eq.~(\ref{pDef}) that with the new variable
\begin{equation}
F_x=\frac{1}{\sum\limits_{x=x_{min}}^{1}p_x/x}\frac{p_x}{x}.
\label{FrP}
\end{equation}
Substitution of Eq.~(\ref{pofs}) to Eq.(\ref{FrP}) yields for
$\langle x \rangle \ll1$
\begin{equation}
F_x=\frac{1}{E_1(x_{min}/\langle x \rangle)}x^{-1}e^{-(x-x_{min})/\langle x\rangle},
\label{FsP}
\end{equation}
where $E_1(z)=\int\limits_{z}^{\infty}e^{-t}/t \ud t$ is the exponential integral.
The distribution given by Eq.(\ref{FsP}) has the form of a power-law
with exponential cutoff.

These simple results highlight the important of parameters like $x_{min}$, $\langle x \rangle$ etc. To get a theoretical handle on these parameters we turn now to a scaling theory that is motivated by
Ref. \cite{75IM}.

\subsection{Domain Size and Magnetic Discontinuities in Amorphous Magnets}

In this section we shall develop scaling arguments for the domain sizes and magnetic discontinuities in amorphous magnets \cite{75IM}. The parameters at our disposal at $T=0$ include the average exchange interaction $\bar J$; the number of magnetic neighbors $q$;  the average magnetic anisotropy strength $\bar K$; the magnetic field $B$; and the system size $N$. For the main body of simulations the values of these parameters were computed in Ref. \cite{13HPS} with the results
\begin{equation}
N=2000,\quad N_A=1300, \quad \bar J\approx .05, \quad q\approx 7, \quad \bar K\approx .08 \ . \label{values}
\end{equation}

\subsubsection{Minimal and Average Domain Size Dependence in the Absence of an Applied Field}

We begin by considering the domain structure in a freshly prepared sample in the absence of a magnetic field. There will be a distribution of domain sizes given by Eq.~(\ref{FsP}). We therefore need to estimate both $x_{min}(\bar J,\bar K)$ and $\langle x\rangle(\bar J,\bar K)$. We do this using estimates for the minimal domain wall energy created by the formation of a domain $E_{w,min}$; the typical domain wall energy created by the formation of a domain $E_{w,typ}$; and the typical anistropy energy $E_{anis}$ for a domain of lengthscale $\xi$.

\noindent As the spins can be rotated in a continuous manner, we find in $d$ dimensions \cite{75IM}:
\begin{equation}
\label{16}
E_{w,min}\sim \bar J\xi^{d-2};
\end{equation}
while
\begin{equation}
\label{17}
E_{w,typ}\sim \bar Jq\xi^{d-2};
\end{equation}
 and
\begin{equation}
\label{18}
E_{anis}\sim -\bar K\xi^{d/2}.
\end{equation}
 Note that the domain wall energy is positive, and each domain of size $s\sim \xi^d$ is created because it can  choose a favorable average orientation $\phi$ for the spins in the domain  which is assumed to be magnetically ordered via the exchange interaction $J$.

To estimate $s_{min}(\bar J,\bar K)$ we now consider the minimal energy cost to create a  domain of size $\xi$. This will be $E_{\xi}\approx E_{w,min}+E_{anis}\sim  \bar J\xi^{d-2}-\bar K\xi^{d/2}$. Thus domains of size $\xi >\xi_{min}$ can exist where
\begin{eqnarray}
\label{19}
\xi_{min} & \sim & (\bar J/\bar K)^{2/(4-d)} \nonumber \\
s_{min}(\bar J,\bar K) &\sim & \xi_{min}^d  \sim  (\bar J/\bar K)^{2d/(4-d)} .
\end{eqnarray}

Specifically in two dimensions $s_{min}(\bar J,\bar K)  \sim  (\bar J/\bar K)^2 $. Using the values shown in Eq. (\ref{values}) we find that $s_{min} \sim O(1)$. As a consequence $\Delta m_{min}= s_{min}/N_A \sim O(10^{-3})$. The simulations have been performed in a region of parameter space where the random anisotropy is strong. The reader should compare this value of $\Delta m$ to the numerically used value $\Delta m_{min}=0.002$.

Let us now estimate $\langle s\rangle(\bar J,\bar K)$. Scaling arguments would suggest that here we need to equate the magnitude of the typical domain wall energy cost to create a  domain of size $\langle \xi \rangle$ to the magnitude of the anisotropy energy, or $ \bar Jq\langle \xi \rangle ^{d-2} \sim \bar K\langle \xi \rangle^{d/2}$ . We thus find
\begin{eqnarray}
\label{20}
\langle \xi \rangle & \sim & (\bar Jq/\bar K)^{2/(4-d)} \nonumber \\
\langle s \rangle &\sim & \langle \xi \rangle ^d  \sim  (\bar Jq/\bar K)^{2d/(4-d)} .
\end{eqnarray}

Specifically in two dimensions $\langle s \rangle  \sim  (\bar Jq/\bar K)^2 $. Using the simulation values therefore $\langle s \rangle  \approx 100 $ and $\langle \Delta M\rangle \sim \langle s \rangle/N  \approx 0.1 $. Another consequence of our estimates Eqs.~(\ref{19}) and ~(\ref{20}) is that $\langle s \rangle/s_{min} \approx q^{2d/(4-d)}$. Thus in and $d=2$ $\langle s \rangle/s_{min} \approx 50$. This justifies the approximation made at the end of Appendix \ref{maxent}. The reader should note that these estimates are strong functions of the values of the parameters; if for example $\bar K$ were reduced for a fixed value of $\bar J$, the domain sizes would increase accordingly.
\subsubsection{Hysteresis Curve and Magnetic Domain Flips in Applied Fields}

\begin{figure}
\includegraphics[scale = 0.26]{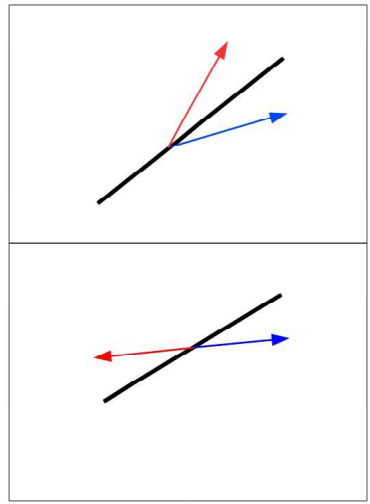}
\caption{Two possible flips of the average spin orientation $\phi$ when the magnetic field ramps up. The present $\phi$ (in red) can jump to $2\theta-\phi$ (in blue), upper panel. A second flip can happen as shown in the lower panel i.e. $\phi\to \phi+\pi$. }
\label{spin}
\end{figure}

Let us now consider the effects of an applied field $B$ on the amorphous magnetic solid. As the field $B$ is cycled a series of distinct irreversible magnetic domain flips followed by reversible domains re-orientation, mapping out the observed hysteresis loop. We described above the domain structure initially when $B=0$. It will consist of domains oriented equally between $-\pi < \phi <\pi$.  As $B$ is now increased (and assumed pointing along the positive x axis), there will be three types of domain flips.
The type of flip responsible for the largest possible $\Delta m$ occurs when $\phi$ flips to $\phi=0$ in one go. We will refer to such flips as type 1. Smaller values of $\Delta m$ will occur upon flips of domains with average spin orientation $\phi$ from $\phi \rightarrow 2\theta-\phi$ provided $\phi <\pi/2$ or $\phi >-\pi/2$, see Fig. \ref{spin}.
The last type of flip is $\phi\to \phi+\pi$, see Fig. \ref{spin}.
 These last two flips are referred to as flips of type 2 and 3 respectively. We will argue in the next subsection that the last two flips contribute on the average (over $\theta$) the same order of magnetization changes.
Such flips cost very little energy, because the anisotropy energy does not change and neither does the exchange interaction. There is only a small domain wall energy that will need to be overcome in order to reduce the magnetic energy. Thus for such a flip to occur
\begin{equation}
\label{a}
E_{f}\sim  -2B \xi^d \cos{\phi}+\bar J\xi^{d-2}< 0,
\end{equation}
or for $d=2$ the domain size must be greater than
\begin{equation}
\label{b}
s > \bar J/2B \cos{\phi}.
\end{equation}
Thus in principle any large enough domain will flip. As, however, their sizes follow the distribution (\ref{pofs}), the typical size of flip will involve domains of size $\langle s \rangle\sim (\bar Jq/\bar K)^2$ and therefore the applied field will need to be of size
\begin{equation}
\label{c}
B_{f}\sim \bar K^2/(\bar J q^2).
\end{equation}
Using the values (\ref{values}) we see that the typical magnetic field $B$ required for these flips is of the order $B_{f}\approx .0025$. These domain flips will be observed as magnetization discontinuities in the $B-M$ hysteresis curve at low magnetic fields.

At larger magnetic fields one can begin to observe flips of type 1, i.e. flips of the form $\phi\rightarrow 0$. Such flips cost more energy, because the contribution of the anisotropy to the energy will change, though the exchange interaction does not. Thus for such a flip to occur
\begin{equation}
\label{a}
\tilde E_{f}\sim  -B \xi^d (1-\cos{\phi})+\bar K\xi^{d/2}< 0,
\end{equation}
or for $d=2$ the domain size must be greater than
\begin{equation}
\label{b}
\tilde s > (\bar K/B (1-\cos{\phi}))^2.
\end{equation}
Again in principle any large enough domain will flip. But, again because the domain sizes follow the distribution (\ref{pofs}), the typical size of $\phi\rightarrow 0$ flips will involve domains of size $\langle s \rangle\sim (\bar Jq/\bar K)^2$ and therefore the applied field will need to be of size
\begin{equation}
\label{c}
\tilde B_{f}\sim \bar K^2/(\bar J q).
\end{equation}
For our values of the parameters we see that the typical magnetic fields $B$ required for these $\phi\rightarrow 0$ transitions to occur are of magnitude $B_{flip,2}\approx .018$. These domain flips will be observed as magnetization discontinuities in the $B-M$ hysteresis curve at larger applied fields.

The important and unavoidable consequence of these scaling arguments is that the pdf
of Barkhausen noise is not homogeneous along the hysteresis curve; it can
change simple because the typical magnitude of observed flips are different at different
values of the magnetic field. It is important to respect this insight when we estimate the Barkhausen statistics.

\subsection{Barkhausen Statistics for a fresh sample}

When we start to ramp the field of a freshly prepared glass, we have the simplification that the orientations of the magnetic domains are
random in the interval $[0,2\pi]$. This is not the case on the hysteresis curve as explained below. We thus start with this simpler case.

The orientation of each magnetic domain can be parameterized by an angle $\phi$, which is the average orientation of the spins in the domain. When the magnetic field is zero, we expect that the angle $\phi$ will not be too far from the local easy axis $\theta$, which is the average of $\theta(\B r_i)$ over the domain.
 \subsubsection{The pdf for large flips}

 The simplest calculation is for the larger magnetic flips for which we assume that changing the magnetic field $B$ results in a giant flip of a whole domain such that $\phi\to 0$. Below we will consider also the smaller flips in Fig.~\ref{spin} and argue that the final result is {\em not the same}. In the present case  the magnetic jump $\Delta M$ will be of size
\begin{equation}
\label{10}
\Delta m = x(1-\cos{\phi}) \ .
\end{equation}
\begin{figure}
\includegraphics[scale = 0.40]{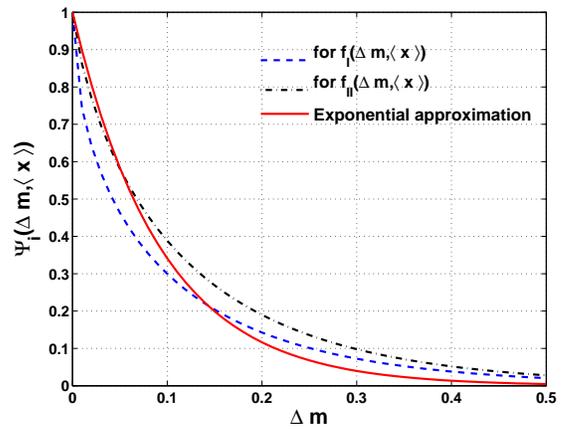}
\caption{Cutoff functions estimated with $\langle x\rangle=0.093$.}
\label{CutOff}
\end{figure}
As a first step we find the conditional probability
$P(\Delta m|x)$ using the fact that
\begin{equation}
\label{11}
P(\phi)  \approx  \frac{1}{2\pi}.
\end{equation}
We note that this last estimate will be correct also during the evolution of the magnetization curve because we deal with domains that did not flip yet.  Then we can write
\begin{equation}
\label{12}
P(\Delta m|x)= \int_{-\pi}^{\pi}\frac{d\phi}{2\pi}\delta (\Delta m-x(1-\cos{\phi})).
\end{equation}
An immediate calculation yields
\begin{equation}
\label{13}
\delta(\Delta m -x(1-\cos{\phi}))=\frac{\delta (\phi-\phi_0)}{x\sin\phi_0} \ ,
\end{equation}
where
\begin{equation}
\cos \phi_0 = 1-\Delta m/x \ , \quad  \text{for}~ x\ge \Delta m/2 \ .
\label{cos}
\end{equation}

At this point we should also average over $x$ to get $P(\Delta m)$:
\begin{equation}
\label{131}
P(\Delta m)=\frac{C}{2\pi} \int_{\Delta m/2}^{1}dx \frac{F_x}{ x \sin{\phi_0}}.
\end{equation}
where $C$ is the normalization constant defined by the condition
$\int\limits_{\Delta m_{min}}^{1}P(\Delta m) \ud \Delta m=1$.

Now from Eq. (\ref{cos}) we find
\begin{equation}
\sin{\phi_0}=\frac{\Delta m}{x}\sqrt{2\frac{x}{\Delta m}-1}\ .
\end{equation}

and therefore Eq.~\ref{131} can be rewritten
\begin{equation}
P(\Delta m)=C\frac{1}{2\pi }\int_0^{\sqrt{2/\Delta m-1}}dz F_{[\Delta m/2(z^2+1)]} \ .
\end{equation}

Using the explicit form of $F_x$ we reach the final result
\begin{equation}
\label{15}
P(\Delta m)=C\frac{\Psi_{cut}(\Delta m,\langle x \rangle) }{\Delta m} \ ,
\end{equation}
where the cutoff function is defined by
\begin{equation}
\Psi_{cut}(\Delta m,\langle x \rangle)=exp(-\Delta m/2\langle x\rangle)f_I(\Delta m,\langle x \rangle,)
\label{151}
\end{equation}
and
\begin{equation}
f_I(\Delta m,\langle x \rangle)=\frac{2}{\pi } \int_0^{\sqrt{2/\Delta m-1}}dz \frac{\exp[-(\Delta m/2\langle x \rangle) z^2]} {z^2+1}
\label{FunI}
\end{equation}

The reader should note that the analytic result given by Eq.~(\ref{15})
is close to the form found numerically, cf. Fig. \ref{comcum}, with exponent
$\alpha=-1$ and an exponential cutoff but with an additional correction in the
form of $f_I(\Delta m)$. It follows from the numerical analysis that for the
value of $\langle x \rangle$ estimated from simulations the upper limit
of the integral in Eq.~(\ref{FunI}) can be replaced by infinity. In this
case the integral can be found in close form and the cutoff function is
given by
\begin{equation}
\Psi_{cut}(\Delta m,\langle x \rangle)=erfc\bigg(\frac{1}{2}\sqrt{2\Delta m/\langle x \rangle}\bigg),
\label{CutIan}
\end{equation}
where $erfc(x)=(2/\sqrt{\pi})\int\limits_{x}^{\infty}ext(-t^2)\ud t$ is the complimentary error function. A comparison of this cutoff function with the exponential approximation for the cutoff is shown in Fig.~\ref{CutOff}.

It now becomes obvious why it is difficult to distinguish, using numerics only, between the approximate form of a power-law time and exponent and the actual statistics found here. We also note that it is not allowed to perform too much asymptotics. We could
advance as follows: an asymptotic expansion of
the complimentary error function is given by $erfc(x)\sim exp(-x^2)/x$,
therefore, for $\Delta m\gg \langle x \rangle$ Eq.~(\ref{15}) is reduced to
a widely advertised form
\begin{equation}
P(\Delta m)\sim\frac{exp(-\Delta m/(2\langle x\rangle))}{\Delta m^{3/2}}.
\label{F1.5}
\end{equation}
However we warn the reader that the limit $\Delta m\gg \langle x \rangle$ does not exist in our theory, and therefore this step is illegal.

\subsubsection{The pdf for somewhat smaller flips}

At smaller values of $B$ the prevalent flips are of types 2 and 3 as shown in Fig.~\ref{spin}. The change in $\Delta m$ in these cases is
\begin{eqnarray}
\Delta m &=& x (\cos(2 \theta -\phi)-\cos \phi) \ , \quad \text{flip 2}\ ,\label{dm1}\\
\Delta m &= &-2x \cos \phi \ , \quad \text{flip 3} \ . \label{dm2}
\end{eqnarray}
In fact, these two flip result in exactly the same theory, since for flip 2 we need to first average over all orientations $\theta$. It is easy to see that the result of this integration leads again to Eq. (\ref{dm2}). The subsequent calculation differs from the previous subsection only in replacing Eq. (\ref{cos}) by
\begin{equation}
\cos \phi_0 = \Delta m/2x \ , \quad  \text{for}~ x\ge \Delta m/2 \ .
\label{cos2}
\end{equation}
Continuing as before one ends up with the pdf in the form of Eq.~(\ref{15})
with the cutoff fonction $\Psi_{cut}(\Delta m,\langle x \rangle)=exp(-\Delta m/2\langle x\rangle)f_{II}(\Delta m,\langle x \rangle)$
 where
\begin{equation}
f_{II}(\Delta m)=\frac{2}{\pi} \int_0^{\sqrt{2/\Delta m-1}}dz \frac{\exp[-(\Delta m/2\langle x \rangle) z^2]} {(z^2+1)\sqrt{z^2+2}}
\end{equation}
This function can be evaluated only numerically, the result is shown in
Fig.~\ref{CutOff}.
\subsection{The pdf along the hysteresis curve}

The calculation of the pdf of magnetic jumps along the hysteresis curve requires a further discussion. It is seen very clearly that upon returning from saturation with $m=1$ the magnetization curve is essentially smooth until the magnetic field changes sign. The reason for this is that the increase in magnetic field beyond saturation forced all the spins to point in the direction of $\B B$. Upon decreasing $\B B$ the values of $\phi$ will return to their positions closer to $\theta$ in every domain, but will not begin to flip before $\B B$ changed signs. Remembering that from the point of view of the local anisotropy term alone there are four energetically equivalent positions of $\phi$ with respect to $\theta$, it is obvious that the smooth relaxation curve will not return $\phi$ to be distributed in the interval $[-\pi,\pi]$, but only to the interval $[-\pi/2,\pi/2]$;  there is no reason to flip direction before $\B B$ changes sign.

In terms of the calculation of the pdf of magnetization jumps all that this amounts to is a change in the limits of integration in equations like (\ref{12}), but this is irrelevant due to the existence of the $\delta$-function. We thus conclude that the pdf in the freshly quenched system and in the hysteresis loop are the same once we excluded very small jumps that may occur along the smooth parts of the hysteresis curve.

\section{Different Parameters}
For the sake of completeness we discuss briefly the effect of changing the parameters on the Barkhausen statistics.
\begin{figure}
\includegraphics[scale = 0.40]{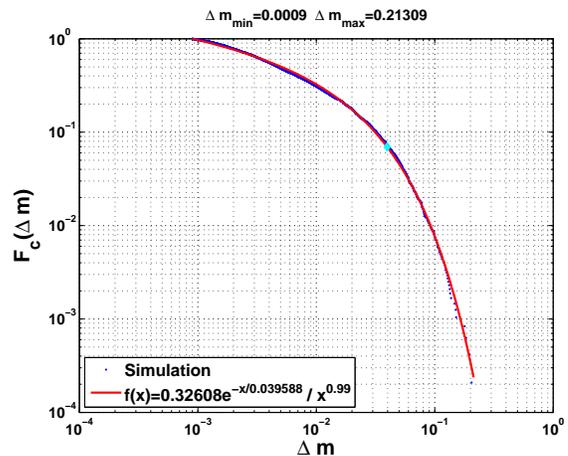}
\caption{The effect of changing the parameters on the Barkhausen statistics. Here we doubled the value of $K$ without changing $J$.}
\label{2K}
\end{figure}
We have doubled the value of $K$, keeping $J$ constant. One expects that this will result in {\em smaller} domains, and therefore in a smaller $\Delta m_{\rm min}$. Accordingly also the values of $B$ where flips occur will change, but nevertheless we do not expect much change in the theory. All these expectations are validated by the results, see Fig. \ref{2K}. A similar misleading power times exponential cutoff is indicated by the data, but we now know that the correct values is $\alpha=-1$ and the apparent exponent 0.99 is spurious. In fact for the present values of parameters the apparent exponent is almost exact. Nevertheless the cutoff functions are not pure exponentials as shown in the previous sections. We thus conclude that at least under a change in parameters the theoretical pdf remains invariant except for a strong renormalization in the range of validity in terms of the minimum and maximum values of $\Delta m$.

\section{Summary and discussion}
\label{summary}

In summary, we believe that we have presented a rather complete theory of Barkhausen statistics in a magnetic glassy model which has a very good chance to represent Barkhausen Noise in metallic glasses. While we claim no universality, we have identified the Barkhausen Noise as resulting from plastic instabilities that occur while the magnetic field is ramped up or down. Simultaneous with the magnetization jumps we have
also energy and stress discontinuities. The statistics of the phenomenon is delicate.
It is not uniform during the ramping of the magnetic field, since the type of flips
changes, from relatively smaller flips when they start at a relatively small value
of the magnetic field $B$ to relatively larger flips at larger values of $B$. We presented a careful theory of the pdf of the magnetization jumps in both regions,
and they have the form shown in Eq. (\ref{15}) and (\ref{151}). Thus, besides
having a power law with $\alpha=-1$ and an exponential cutoff, we have an additional
cutoff function, $f_I$ or $f_{II}$ which are effectively changing the power law if not properly identified. They are the reason for the apparent exponent $\alpha=-1.05$ in Fig.~\ref{comcum} lower panel.

\acknowledgments

This work had been supported in part by an ERC ``ideas" grant STANPAS, the Israel Science Foundation
and by the German Israeli Foundation.

\appendix
\section{Maximum Likelihood Method}
\label{maxlike}
The maximum likelihood estimation introduced in \cite{F12} (see, also, \cite{F70}) aims at estimating
of the parameters of a statistical model. A parametric statistical model in the case of one random
variable $x$ is defined by the probability density function $p(x\mid{\B \alpha})$,
where ${\B \alpha}=\{\alpha_i\}$ is a set of parameters of the model. For a random sample ${\B X}=\{X_i\}$
($1\le i\le N$) of  {\it independent and identically distributed} observed values the joint probability density function is given by
\begin{equation}
{\bar p}({\bf X}\mid{\bf \alpha})=\prod\limits_i^N p(X_i\mid{\bf \alpha}).
\label{A1}
\end{equation}
For a given sample the set ${\bf X}$ can be considered as fix parameters of the function defined by
Eq.~(\ref{A1}) and ${\bf \alpha}$ are the function's free varying variables.   Therefore, under these conditions,
the likelihood function is defined as
\begin{equation}
{\bf L}({\bf \alpha}\mid {\bf X})={\bar p}({\bf X}\mid {\bf \alpha}).
\label{A2}
\end{equation}
It is often more convenient to use logarithm of the likelihood function called the log-likelihood function
\begin{equation}
{\bf l}({\bf \alpha}\mid {\bf X})=\sum\limits_i^N \ln p(X_i\mid {\bf \alpha})
\label{A3}
\end{equation}
The method of maximum likelihood estimation consists in finding values of the set ${\bf \hat{\alpha}}$ that
maximized this function
\begin{equation}
{\bf \hat{\alpha}}=\underset{{\bf \alpha}}{\arg\max}\hspace{1mm}{\bf l}({\bf \alpha}\mid {\bf X}).
\label{A4}
\end{equation}
If the log-likelihood function is differentiable and ${\alpha_i}$ exist its maximum is defined by
a solution of  likelihood equations
\begin{equation}
\frac{\partial {\bf l}({\bf \alpha}\mid {\bf X})}{\partial \alpha_i}=0.
\label{A5}
\end{equation}
In the case when a  statistical model involves many parameters and its probability density function  is highly non-linear  the solution of Eq.~(\ref{A4}) can be find with optimization algorithms. A simplest way consists in
the evaluation of the log-likelihood function on a grid in a space of parameters $\alpha_i$.
The power law distribution with an exponential cutoff is defined by
\begin{equation}
p(x\mid x_{min},\gamma,x_0)=\frac{x^{-\gamma}e^{-x/x_0}}{\int_{x_{min}}^{\infty}x^{-\gamma}e^{-x/x_0}\bm{d} x},x\ge x_{min}.
\label{A6}
\end{equation}
Substitution of Eq.~(\ref{A6}) to Eq.~(\ref{A3}) yields the log-likelihood function in this case
\begin{eqnarray}
{\bf l}(x_{min},\gamma,x_0 \mid {\bf X})&=&-\gamma\sum\limits_{i}^{N}\ln X_i -\frac{1}{x_0}\sum\limits_{i}^{N} X_i\nonumber \\
&-&\ln \int_{x_{min}}^{\infty}x^{-\gamma}e^{-x/x_0}\bm{d} x
\label{A7}
\end{eqnarray}

\section{Maximum entropy}
\label{maxent}
Let Lagrangian is
defined by
\begin{eqnarray}
L&=&-\sum\limits_{s=s_{min}}^{N} p_s\ln p_s-\lambda_1\bigg( \sum\limits_{s=s_{min}}^{N} p_s-1\bigg) \nonumber\\&-&\lambda_2\bigg(\sum\limits_{s=s_{min}}^{N} s p_s-\langle s\rangle\bigg),
\label{lagr}
\end{eqnarray}
where $\lambda_1$ and $\lambda_2$ are Lagrange multipliers.
Setting the partial derivatives of Eq.~(\ref{lagr}) with respect to $p_s$
to zero yields
\begin{equation}
-\ln p_s-1-\lambda_1-\lambda_2 s=0.
\label{Leq}
\end{equation}
Solution of this equation reads
\begin{equation}
p_s=e^{-\lambda_1-1-\lambda_2 s}.
\label{sol1}
\end{equation}
Substitution of Eq.~(\ref{sol1}) to Eq.~(\ref{norm}) yields
\begin{equation}
e^{-\lambda_1-1}=\frac{1}{Z(\lambda_2 )},
\label{norm1}
\end{equation}
where $Z(\lambda_2 )=\sum\limits_{s=s_{min}}^{N} e^{-\lambda_2 s}$ and the solution
given by Eq.(\ref{sol1}) reads
\begin{equation}
p_s=\frac{1}{Z(\lambda_2 )} e^{-\lambda_2 s}.
\label{sol2}
\end{equation}
Substitution of Eq.~(\ref{sol2}) to Eq.~(\ref{mean}) yields the condition which
defines the constant$\lambda_2$
\begin{equation}
\frac{\partial \ln Z(\lambda_2)}{\partial \lambda_2}=-\langle s\rangle.
\label{eqL}
\end{equation}
In order to evaluate the function $Z(\lambda_2)$ the sum can be replaced by
the integral
\begin{eqnarray}
Z(\lambda_2 )&=&\sum\limits_{s=s_{min}}^{N} e^{-\lambda_2 s}
\approx\int\limits_{s=s_{min}}^{N} e^{-\lambda_2 s}\ud s\nonumber\\
&=&\frac{1}{\lambda_2}\bigg(e^{-\lambda_2 s_{min}}-e^{-\lambda_2 N}\bigg)\nonumber \\
&\to_{N\to \infty}&\frac{e^{-\lambda_2 s_{min}}}{\lambda_2}.
\label{estZ}
\end{eqnarray}
It follows from this equation and Eq.~(\ref{eqL}) that the parameter
$\lambda_2$ is defined by
\begin{equation}
\lambda_2=\frac{1}{\langle s \rangle -s_{min}}.
\label{L2}
\end{equation}
Substitution of this solution to Eq.~(\ref{sol2}) yields the following
approximation of the probability distribution function
\begin{equation}
p_s=\frac{ e^{\frac{s_{min}}{\langle s \rangle -s_{min}}}}{\langle s \rangle -s_{min}} e^{-\frac{s}{\langle s \rangle -s_{min}}}.
\label{pdfP}
\end{equation}
For $s_{min}\ll\langle s\rangle$ this equation is reduced to Eq. (\ref{pofs})

\end{document}